\documentclass[
    ,final            
  ]
  {aipproc}

\layoutstyle{6x9}

\usepackage{amsmath}
\usepackage{amssymb}
\usepackage{graphicx}
\usepackage{multirow}
\newcommand{\bZ}{\mathbb{Z}}

\newcommand{\bR}{\mathbb{R}}
\newcommand{\cN}{\mathcal{N}}

\newcommand{\ov}{\overline}

\begin{document}

\newcommand{\papertitle}{Branes and instantons at angles and the
  F-theory lift of $O(1)$ instantons}
\title[\papertitle]{\papertitle\footnote{Based in part on talks given
    at Supersymmetry'09 (M.C.), Strings'09 (M.C.), CERN (I.G.-E.) and
    Simons Workshop in Mathematics and Physics, Stony Brook (M.C.).}}



\author{M. Cveti{\v c}}{
  address={Department of Physics and Astronomy, University of Pennsylvania,\\
  Philadelphia, PA 19104-6396, USA}
}

\author{I. Garc\'ia-Etxebarria}{
  address={Department of Physics and Astronomy, University of Pennsylvania,\\
  Philadelphia, PA 19104-6396, USA}
}

\author{R. Richter}{
  address={Dipartimento di Fisica, Universit\`a di Roma ``Tor Vergata''\\
    I.N.F.N. - Sezione di Roma ``Tor Vergata''\\
    Via della Ricerca Scientifica, 1 - 00133 Roma, ITALY
}
}

\begin{abstract}
  We discuss the physics of D-branes and D-brane instantons
  intersecting at angles, focusing on the (non)generation of a
  superpotential in the worldvolume theory of the branes. This is a
  short review of the results in arXiv:0905.1694, where we further
  emphasize both the macroscopic and microscopic structure of the
  manifestly supersymmetric instanton action. We also comment on the
  lift of $O(1)$ instantons to F-theory.
\end{abstract}

\begin{flushright}
  ROM2F/2009/21\\
  UPR-1214-T
\end{flushright}

\maketitle

\section{Introduction}

In the last few years there has been growing interest in
non-perturbative effects coming from Euclidean D-brane instantons
\cite{Becker:1995kb}, mainly due to the realization
\cite{Blumenhagen:2006xt,Ibanez:2006da,Florea:2006si} that they could
be used for solving some longstanding difficulties in string model
building coming from perturbative symmetries. We refer the reader to
\cite{Blumenhagen:2009qh} for a review on  many of the recent
developments in this topic.

Since \cite{Blumenhagen:2009qh} appeared there has been some progress
in the topic. In these notes we will review some of the results in
\cite{Cvetic:2009mt}, putting them in a somewhat broader
perspective. In particular, we will discuss how supersymmetry
determines the result almost uniquely, in a manner compatible with
both our global macroscopic arguments and our microscopic CFT
computations. We will also describe some results about the lift to
F-theory of $O(1)$ instantons.

\medskip

We will focus on instantons contribution to the superpotential. In
order for an instanton to contribute to the superpotential, it must
have exactly two unsaturated fermionic zero modes, corresponding to
the two $\theta^\alpha$ $\cN=1$ superspace fermionic coordinates
($\alpha\in\{1,2\}$) over which we integrate the superpotential.

A generic rigid D-brane instanton in a type II background preserving 8
supersymmetries will have at least 4 fermionic zero modes living on
its worldvolume, coming from the broken background supersymmetries. We
will call these modes $\theta^\alpha$ and
$\ov\tau_{\dot\alpha}$. These zero modes are the Goldstinos of the
supersymmetries broken by the D-brane (which is a 1/2 BPS state). Only
those instantons for which the $\ov\tau$ modes are projected out or lifted in
going from $\cN=2$ to $\cN=1$ can contribute to the superpotential.

There are various known ways in which the $\ov\tau$ modes can get
lifted in a realistic $\cN=1$ compactification. The simplest class of
instantons correspond to the lift to string theory of gauge theory
instantons. They are given by D-brane instantons wrapping the same
cycle as a background gauge brane\footnote{More precisely, they must
  wrap the same cycle and have identical worldvolume bundle. If the
  worldvolume bundle is different the instanton is not an ordinary
  field theory gauge instanton, and as we will show below it does not
  contribute to the superpotential.} \cite{Witten:1995gx,Douglas:1996uz,Billo:2002hm,Akerblom:2006hx,Bianchi:2007fx,Argurio:2007vqa,Bianchi:2007wy}. The case with a
single space-filling brane (the so-called $U(1)$ instanton) is
slightly special, since it does not admit a gauge theory
interpretation, but the lifting of $\ov\tau$ modes still takes place
\cite{Aganagic:2007py,GarciaEtxebarria:2007zv,Petersson:2007sc,GarciaEtxebarria:2008iw,Ferretti:2009tz}.

Another class of contributing instantons are those which have a
worldvolume $O(1)$ gauge symmetry, due to being invariant under an
appropriate background orientifold. The orientifold will project out
the $\ov\tau$ modes
\cite{Argurio:2007qk,Ibanez:2007rs,Argurio:2007vqa,Bianchi:2007wy}.

In a realistic compactification we will also need to introduce
background fluxes. These fluxes have a potentially dramatic effect on
the instanton, and in some cases they can also lift the $\ov\tau$
modes
\cite{Marolf:2003ye,Marolf:2003vf,Kallosh:2005gs,Martucci:2005rb,Bergshoeff:2005yp,Park:2005hj,Bandos:2005ww,Lust:2005cu,Tsimpis:2007sx,Schulgin:2007zz,Blumenhagen:2007bn,GarciaEtxebarria:2008pi,Billo':2008pg,Billo':2008sp,Uranga:2008nh}.
This is generically hard to do without breaking supersymmetry
\cite{Blumenhagen:2007bn,GarciaEtxebarria:2008pi,Billo':2008pg,Billo':2008sp,Uranga:2008nh},
for reasons best understood from the effective field theory analysis
\cite{GarciaEtxebarria:2008pi,Uranga:2008nh}. In \cite{Cvetic:2009mt}
we introduced a configuration, connected to $\cN=4$ instantons in
their Coulomb branch, in which flux lifts $\ov\tau$ modes, and which
can be studied using open string techniques. We will review it below.

So far these are the only known configurations in which a D-brane
instanton can contribute to the low-energy superpotential. It is an
interesting open problem to determine whether these are all the
possibilities or other configurations exist. Particularly interesting
is the case where background branes intersect the instantons at
angles, since this is a very generic occurrence in string
compactifications. If the brane intersecting the instanton could also
lift the $\ov\tau$ modes of the instanton this would allow for great
flexibility in model building. As an example, related configurations
where explored in \cite{Heckman:2008es,Marsano:2008py} as a possible
source of interesting non-perturbative dynamics in local F-theory
models.

In order to determine the fate of this class of systems, in
\cite{Cvetic:2009mt} we performed a careful analysis of a IIA system
composed of a D6 brane and a rigid D2-brane instanton intersecting at
angles in the internal space in the absence of orientifolds or fluxes,
which we summarize here. Our main result is that in this case the
$\ov\tau$ modes are not lifted, and thus no
superpotential is generated by the D-brane instantons.\footnote{Of
  course, some other mechanism such as fluxes or orientifolds can lift
  the zero modes of the instanton in our setup, allowing it to
  contribute to the superpotential. What we show here is that such a
  lifting cannot come from the background brane intersecting the
  instanton at an angle.} Again, this negative result has important
implications for string theory model building, since generic
configurations cannot generate certain desired interactions, such as
Polonyi-type couplings.

The main technical tool we will use in our analysis is a microscopic
CFT calculation, although we will also discuss some general effective
field theory arguments forbidding the generation of the
superpotential. This is the main reason why we have chosen to work in
type IIA: close to the point(s) where the instanton and the brane
intersect one has an explicit conformal field theory (CFT) description
of the system, and the fact that $\ov\tau$ modes are not lifted
can be shown in detail.

We will be able to extract some lessons from the CFT analysis that we
expect to be of general applicability. The most important one concerns
the supersymmetry of the system. As we will see, obtaining a
supersymmetric instanton action is a very constraining requirement,
which our CFT results satisfy in a non-trivial manner. Related to
this, we will find that in the intersecting case the spectrum of
states is markedly different from the one we obtain in the gauge
theory case, so one should exercise care in generalizing expressions
obtained from gauge theory instantons.

The type IIA intersecting brane picture is related by duality to many
interesting string backgrounds, so our result has a wide range of
applicability. We refer the reader to \cite{Cvetic:2009mt} for details
and references on some known dual configurations where our results
have important consequences.

\section{CFT analysis of instantons at angles}

\subsection{General structure of the instanton action}

Before going into the detailed CFT analysis, let us highlight some
general features of our analysis.

The first important result we obtain from the CFT analysis concerns
the spectrum of massive modes, shown in table~\ref{table spectrum
  instanton}. We will find that in addition to the modes that appear
in the gauge instanton limit, there exist some \emph{additional modes}
(which we will denote below by primes) that only appear in the
intersecting case, and have no analog in the gauge instanton
case. They are required in order to give a mass to the modes that
become massless in the gauge instanton limit. One essential feature is
that $\mu$ modes pair only with $\ov\mu'$ modes, not with $\ov\mu$
modes, as it would be naively expected. This is straightforward to see
from our calculations below, since the mass of $\mu$ becomes different
from the mass of $\ov\mu$ under small non-supersymmetric rotations of
the system.

Furthermore, we will find that the \emph{supersymmetry
  transformations} shown in equation~\eqref{eq: Susy transfo} play a
crucial role in constraining the action. Looking at these
supersymmetry transformations it is easy to see that the mass
couplings for the $\mu$ modes have to be of the form $\mu\ov\mu'$,
agreeing with the general result based on the expressions for the
masses.

More important than the mass couplings, though, are the \emph{cubic
  couplings} in the effective action, shown in
equation~\eqref{eq:coupling-tau}. Whether or not we saturate the
$\ov\tau$ modes will depend on the form of these couplings. The
couplings given in \eqref{eq:coupling-tau} are obtained using CFT, but
agree beautifully with the \emph{requirement of supersymmetry}. It is
easy to see that any other possible coupling one might consider
($\ov\tau^{\dot\alpha}\ov\omega_{\dot\alpha}\mu'$ is one example one
might be tempted to write) will break supersymmetry. More generally,
supersymmetry and the form of the coupling to the auxiliary field
$\vec{D}$ impose that the cubic coupling must be of the schematic
form:
\begin{equation}
  \ov\tau_{\dot\alpha}\delta_{\xi_{\dot\alpha}}(\ov\omega\omega)
\end{equation}
In particular, only fields that appear after supersymmetry variations
of $\omega$ or $\ov\omega$ can couple to $\ov\tau$. In particular this
excludes $\ov\mu'$, the mass partner of $\mu$.

This last property is extremely important and robust, and we believe
that ultimately it is this fact that underlies the null result that we
find (at least microscopically, see below for the macroscopic
effective field theory justification). A similar supersymmetry
analysis is also possible in related systems that do not admit a
direct CFT description, as long as we have some way of determining the
action of supersymmetry on the spectrum of instanton
modes.\footnote{More precisely, for all compactifications with $(0,2)$
  worldsheet supersymmetry the $U(1)$ worldsheet charge is inherently
  related to the space-time supercharge \cite{Banks:1987cy,Banks:1988yz}. Using this fact, one can demonstrate that the $U(1)$
  charge ensures a unique target space supersymmetric structure of the
  instanton action, given by the structure that we obtain in
  equations~\eqref{eq:coupling-mass} and \eqref{eq:coupling-tau}.}

\subsection{Detailed analysis}

Let us proceed to the microscopic analysis. In type IIA the relevant
instantons are Euclidean D2 branes wrapping a rigid three-cycle
$\Sigma$ away from the orientifold plane in the internal
manifold. Generically, such an isolated $U(1)$ instanton exhibits four
bosonic zero modes $x^{\mu}$ corresponding to the breakdown of
four-dimensional Poincar{\'e} invariance and various Goldstino modes
$\theta^{\alpha}_A$ and $\ov \tau^{\dot{\alpha}}_A$ associated with
the background supersymmetries broken by the instanton. The $A$
subindex comes from flat space, and runs from 1 to 4. As explained
above, we will identify one of the flat space supersymmetries as the
restriction to the intersection of the curved space
supersymmetry. This naturally singles out one of the $A$, which we
will take  in our analysis to be $A=0$. We will return to this point below.

The space-time filling D6-brane wraps a cycle $\Sigma'$ distinct from
$\Sigma$, and it will intersect $\Sigma$ at isolated points in the
internal space. Thus in addition to the neutral instanton modes there
are also charged modes which correspond to strings stretching between
the E2-instanton and the D6-brane.

The different massive and massless charged instanton modes appear as
excitations of the NS- and R-vacuum. Schematically they take the form
\begin{equation}
\sum_{k \epsilon \mathbf{Z}} \left(\alpha^I_{k-\theta_I}\right)^{N^I_k}
\left(\psi^I_{k+\frac{1}{2}-\theta_I}\right)^{M^I_k} \,
|0\rangle_{NS} \qquad  \qquad\sum_{k \epsilon
\mathbf{Z}}\left(\alpha^I_{k-\theta_I}\right)^{N^I_k}
\left(\psi^I_{k-\theta_I}\right)^{M^I_k} \, |0\rangle_{R}\,\,,
\label{eq:states}
\end{equation}
where $\theta_I$ is the intersection angle between the D-brane and the
instanton in the $I$-th dimension (we take $\theta_I$ to be defined in
the range $[-1,1)$) and $\alpha^I_k$ and $\psi^I_k$ denote the bosonic
and fermionic creation operators. Note that the fermionic creation
operator $\psi^I_k$ has Fermi statistics, and thus $M^I_k$ can only
take the values $0$ or $1$, while $N^I_{k}$ can be any non-negative
integer.

The mass of these states is given by:
\begin{align}
\text{NS-sector:}&\qquad \qquad  M^2 \sim \varepsilon^{NS}_0 +
\sum_{k}\Big( N^I_k (k-\theta_I) + M^I_k
(k+\frac{1}{2}-\theta_I)\Big)
\label{eq:bosonic-states}
\\
\text{R-sector:}&\qquad \qquad  M^2 \sim \varepsilon^R_0 +
\sum_{k}\Big( N^I_k (k-\theta_I) + M^I_k (k-\theta_I)\Big)\,\,,
\end{align}
where $\varepsilon_0^{NS,R}$ denotes the zero point energy in the NS
or R sectors, and crucially depends on the intersection angles of the
D6-brane and the E2-instanton. Note that due to the Dirichlet-Neumann
boundary conditions in spacetime the zero point energy for the NS
sector is shifted by $1/2$ compared to the D6-D6 brane system. This
implies that there are no tachyonic modes between the D-brane and the
instanton and for non-trivial angles the bosonic modes are always
massive. The absence of tachyons in this setup rules out the
possibility of recombination of the instanton and D-brane. For the
R-sector the Dirichlet-Neumann boundary conditions in spacetime imply
that only states whose $U(1)$-wordsheet charge is $Q_{WS}=
-\frac{1}{2} \,\, \text{mod} \,\,2$ are allowed, thus only chiral
modes survive the GSO-projection
\cite{Blumenhagen:2006xt,Ibanez:2006da,Florea:2006si,
  Cvetic:2007ku}. The latter is related to the fact that the total
fermion number in the R-sector is constrained due to the
GSO-projection.

For concreteness we assume the intersection pattern
\begin{equation}
\theta_1>0 \qquad \theta_2>0 \qquad \theta_3<0 \label{eq:setup}
\end{equation}
and look at all states whose mass squared is at most of linear order
in the intersecting angles $\theta_I$. In table \ref{table spectrum
  instanton} we present all states and their corresponding mass. When
the D-instanton D-brane system preserves the supercharge $\ov
Q^A_{\dot{\alpha}}$ the charged instanton modes are related to each
other in the following way \cite{Cvetic:2009mt}
\begin{align}
\delta_{\ov \xi^{A}} \mu'_A = i\,\ov \xi^A_{\dot{\alpha}} \omega^{\dot{\alpha}} \,\,\,\,\,\,\,\,\,\,\,\,\,\,\delta_{\ov
\xi^{A}} \omega_{\dot{\alpha}}= i \ov \xi^A_{\dot{\alpha}} \mu_A \,\,\,\,\,\,\,\,\,\,\,\,\,\,\delta_{\ov \xi^{A}} \ov\mu'_A = i\,\ov \xi^A_{\dot{\alpha}} \ov\omega^{\dot{\alpha}} \,\,\,\,\,\,\,\,\,\,\,\,\,\, \delta_{\ov
    \xi^{A}} \ov\omega_{\dot{\alpha}}= i \ov \xi^A_{\dot{\alpha}} \ov\mu_A\,\,.
    \label{eq: Susy transfo}
\end{align}
If $\mu_A$ or $\ov \mu_A$ is massless we expect the absence of $\ov
\mu'_A$.  This is analogous to the D6-D6 brane configuration, where we
have in the massive case a hypermultiplet (consisting of two chiral
multiplets) while in the massless case we only have a single chiral
multiplet. In case the D-instanton and D-brane wrap the same cycle all
supercharges $\ov Q^A_{\dot\alpha}$ are preserved and all $\mu_A$ and
$\ov \mu_A$ are massless. Then the SUSY transformations reduce to the
second and fourth equations of \eqref{eq: Susy transfo} which are
exactly the usual SUSY transformations appearing in the ADHM case
\cite{Green:2000ke,Billo:2002hm}\footnote{Note that for the gauge
  instanton setup, which corresponds to all three intersection angles
  $\theta_I$ being $0$, the vacuum is defined differently. In that
  case the $\mu'_A$ and $\ov \mu'_A$ modes are projected out and do
  not appear in the instanton mode spectrum.}.

\begin{table}[h] \centering
\begin{tabular}{ccccc}
  \hline
  E2-a & CFT state & a-E2 & CFT state & mass$^2$\\
  \hline
  \hline
  $\mu_0$     & $\psi_{-\theta_1}\, \psi_{-\theta_2} | 0
  \rangle^R_{E2a} $  & $\ov \mu'_0$ & $\frac{1}{2} \left( 
  \psi_{-\theta_1} \alpha_{-\theta_2} + \psi_{-\theta_2} \alpha_{-\theta_1}\right)| 0 \rangle^R_{aE2} $ & $\theta_1+ \theta_2$\\
  \hline
  $\mu_1$       &$\psi_{-\theta_1}\, \psi_{\theta_3} | 0
  \rangle^R_{E2a} $  & $\ov \mu'_1$ & $\psi_{\theta_3}
  \alpha_{-\theta_1} | 0 \rangle^R_{aE2}$ & $\theta_1 -\theta_3$ \\
  \hline
  $\mu_2$      &$\psi_{-\theta_2}\, \psi_{\theta_3} | 0 \rangle^R_{E2a}$ & $\ov \mu'_2$  & $\psi_{\theta_3} \alpha_{-\theta_2} | 0 \rangle^R_{aE2}$ &  $\theta_2 -\theta_3$\\
  \hline
  $\mu_3$         & $| 0 \rangle^R_{E2a}$ & & {\bf None} & 0\\
  \hline
  $\mu'_0$         &   $\alpha_{\theta_3} | 0 \rangle^R_{E2a} $ & $\ov \mu_0$& $\psi_{\theta_3} | 0 \rangle^R_{aE2}$ & $-\theta_3 $\\
  \hline
  $\mu'_1$        & $\alpha_{-\theta_2} | 0 \rangle^R_{E2a} $ & $\ov
  \mu_1$ &$\psi_{-\theta_2} | 0 \rangle^R_{aE2}$ & $\theta_2$\\
  \hline
  $\mu'_2$             & $\alpha_{-\theta_1} | 0 \rangle^R_{E2a}$ &
  $\ov \mu_2$ & $\psi_{-\theta_1} | 0 \rangle^R_{aE2}$ & $\theta_1$\\
  \hline
  $\mu'_3$ & $\frac{1}{3} \, \epsilon^{2}_{ijk} \psi_{-\theta_i}
  \psi_{-\theta_j} \alpha_{-\theta_k}   | 0 \rangle^R_{E2a} $  & $\ov
  \mu_3$  &$\psi_{-\theta_1} \psi_{-\theta_2} \psi_{\theta_3} | 0
  \rangle^R_{aE2}$ & $\theta_1 +\theta_2 -\theta_3 $\\
  \hline
  $\omega^{\dot{\alpha}}$ & $| 0 \rangle^{NS}_{E2a}$ & $\ov
  \omega^{\dot{\alpha}}$ & $| 0 \rangle^{NS}_{E2a}$ & $\frac{1}{2}\left(\theta_1 +\theta_2 -\theta_3\right) $\\
  \hline
\end{tabular}

\caption{Instanton D-brane spectrum. We have paired modes by
  mass. Notice that, despite what the notation might suggest, $\mu$
  modes can only pair up with $\ov\mu'$ modes, and
  $\mu'$ modes can only pair up with $\ov\mu$ modes. Notice also the
  chiral spectrum at the massless level, encoded in the fact that
  $\mu_3$ has no mass partner.}
\label{table spectrum instanton}
\end{table}

To each state corresponds a vertex operator and one can explicitly
show \cite{Cvetic:2009mt} the existence of the mass terms
\begin{equation} 
S^{mass}_{E2} =\mu_A\, \ov \mu'_A + \omega \ov \omega +  \mu'_A\, \ov \mu_A \,\,.
\label{eq:coupling-mass}
\end{equation}
In order to simplify the expression we have omitted the exact mass
coefficient, it agrees with the result displayed in table~\ref{table
  spectrum instanton}. Note in particular that in the case where all
SUSY is broken by the E2-D6 system the masses of $\mu_A$ and $\ov
\mu_A$ are different, and thus a mass term $\mu_A \, \ov \mu^A$ is
never allowed. Moreover, in case some supercharge $\ov
Q^A_{\dot{\alpha}}$ is preserved the bosonic mass term $\omega \ov
\omega$ implies the presence of the fermionic mass terms $\mu_A \ov
\mu'_A$ and $\mu'_A \ov \mu_A$ to ensure invariance under the SUSY
transformations \eqref{eq: Susy transfo}.

Furthermore, one can also show that the ADHM-like couplings
\begin{align}
S^{ADHM}_{E2}= \ov \tau^A_{\dot{\alpha}} (\ov \omega^{\dot{\alpha}} \mu_A + \ov \mu_A
\omega^{\dot{\alpha}}) +  i
\vec{D}\cdot
  \omega \,\vec{\sigma}\,\bar\omega
\label{eq:coupling-tau}
\end{align}
survives even for non-trivial intersections, where the latter interaction term describes the coupling of the bosonic modes to the auxiliary field $D^{\mu\nu}=\vec{D} \,
\vec{\sigma}_{\mu\nu}$. Taking into account the supersymmetry transformation for $\ov \tau^A_{\dot \alpha}$
\begin{align}
 \delta_{\ov \xi^A} \ov \tau
= \ov \xi^A\vec{\sigma} \vec{D}
\end{align}
it is easy to see that the interaction term \eqref{eq:coupling-tau} is
supersymmetric if the E2-D6 system preserves $\ov Q^A_{\dot \alpha}$.

Let us now investigate whether or not a $U(1)$ instanton intersecting
with a D-brane does contribute to the superpotential by performing the
path integral over the instanton modes explicitly. Let us assume that
the E2-D6 system preserves the supercharge $\ov Q^0_{\dot \alpha}$,
which translates for our concrete setup \eqref{eq:setup} into the
constraint on the angles $\sum_I \theta_I =0$. Moreover, we assume
that the instantons are wrapping a rigid cycle thus apart from
$\theta^{\alpha}\equiv \theta^\alpha_0$ and $\ov
\tau_{\dot{\alpha}}\equiv \ov\tau_{\dot\alpha}^0$, the rest of the
$\ov\tau^A,\theta_A$ modes are lifted by the holonomy of the
background.

The path integral takes the form
\begin{equation}
\int d^4 x d^2\theta d^2 \ov \tau \prod^3_{A=0} \, d^2 \omega d^2\ov
\omega\, d  \mu'_A\, d \mu_A \prod^2_{B=0} d  \ov \mu'_B \, d\ov
\mu_B \,d \ov \mu_3\,\, e^{-S^{mass}_{E2}-S^{ADHM}_{E2} }\,\,.
\end{equation}
After performing the integration over the $\ov \tau$ modes and
$\mu^0$, $\ov \mu^0$ we are left with
\begin{equation}
\int d^4 x d^2\theta  \prod^3_{A=1} \, d^2 \omega d^2\ov \omega\, d
\mu'_A\, d \mu_A \prod^2_{B=1} d  \ov \mu'_B \, d\ov \mu_B \,d \ov
\mu_3 d\mu'_0 d\ov \mu'_0 \,\,e^{-S^{mass}_{E2}}\,\,,
\end{equation}
where we omit the term $\vec{D}\cdot \omega \,\vec{\sigma}
\,\bar\omega $ which is irrelevant for the analysis. Next we use the
mass terms for saturating the remaining fermionic charged instanton
modes:
\begin{equation}
\int d^4 x d^2\theta  \prod^3_{A=1} \, d^2 \omega d^2\ov \omega\, d
\ov \mu_3 d\mu'_0 d\ov \mu'_0 \,\,e^{-( \ov \mu^0 \,
\mu'_{0} +  \mu^{0} \, \ov \mu'_{0})}\,\,.
\end{equation}
Note that we cannot saturate the $\mu'_0$ and $\ov \mu'_0$ modes since
we used already $\mu_0$ and $\ov \mu_0$ to saturate the universal $\ov
\tau$ modes. Since there is no other way to soak up these modes the
whole path integral vanishes. Thus we conclude a generic rigid $U(1)$
instanton does not contribute to the superpotential.
 
\section{Effective field theory argument}

The previous discussion fits well into the general effective field
theory analysis of
\cite{GarciaEtxebarria:2007zv,GarciaEtxebarria:2008pi}. In particular,
\cite{GarciaEtxebarria:2008pi} argues that instantons which can become
non-BPS (relative to the $\cN=1$ background) as we move in closed
string moduli space do not contribute to the superpotential, but
contribute instead to Beasley-Witten higher F-terms
\cite{Beasley:2004ys,Beasley:2005iu}.

There are various important subtleties that one has to keep in mind
when using this argument. First of all, when applying any misalignment
argument one only needs to consider the subsector of the
compactification putatively responsible for lifting the zero modes of
the instanton. For instance, the example of a $U(1)$ instanton
intersecting a brane discussed above will be generically part of some
given compactification involving orientifold planes (away from the
instanton), fluxes,\footnote{As mentioned in the introduction, fluxes
  can change the analysis in important ways, and we do not address
  them in this section.} and more background branes. In such a
complete compactification, the physical closed string moduli space is
very restricted (or possibly absent altogether, in the case where we
stabilize all moduli), and typically one cannot misalign the instanton
with the background while keeping supersymmetry. Nevertheless, those
ingredients that have demonstrably no effect on $\ov\tau$ lifting can
be ignored. Zero modes will be lifted in the complete compactification
if and only if they are lifted in the absence of these elements of the
compactification. The most important effect this has in practice is
that we can generically ignore orientifold planes away from the
instanton. We only need to address the instanton and D-branes in a
type II Calabi-Yau background.

Once we determine that the background orientifolds do not lift the
zero modes of the instanton, the misalignment argument in
\cite{GarciaEtxebarria:2008pi} becomes very powerful: we can view our
set of branes as a collection of 1/2 BPS states of the $\cN=2$
Calabi-Yau background, with its closed string moduli space
intact. Moving in this closed string moduli space will change the
argument of the central charge of the branes, differently for each
homology class. Let us illustrate this in detail for type IIA
compactifications. In this case the central charge of a brane
(possibly instantonic) wrapping a 3-cycle $\Sigma$ in the internal
Calabi-Yau is:
\begin{equation}
  \label{eq:zentral}
  Z = \int_\Sigma \Omega,
\end{equation}
with $\Omega$ the complex structure of the Calabi-Yau. Type IIB at large volume work similarly, just replace $\Omega$ by
$e^{-J+i(B-F)}$ in equation \eqref{eq:zentral}:\footnote{We have
  omitted some curvature terms that do not modify our discussion, see
  for example \cite{Aspinwall:2004jr} for the complete expression.}
\begin{equation}
  \label{zentral-IIB}
  Z = \int_X \mathrm{Tr}\,e^{-J+i(B-F)}
\end{equation}
with $X$ the cycle wrapped by the brane mirror to $\Sigma$, and $F$
the worldvolume flux on the mirror brane. In particular, note that the
effect of a non-trivial worldvolume flux is to modify the effective
homology class on which the brane is wrapped. At small volume the type
IIB expression receives large $\alpha'$ corrections, and it is more
useful to map the problem to IIA using mirror symmetry.

Which $\cN=1$ supersymmetry is preserved is determined by $\arg(Z)$,
the phase of $Z$. This phase depends only on the homology class
$[\alpha]$ of $\Sigma$, and our location in complex structure moduli
space. Since in the absence of orientifolds we can move in arbitrary
directions in complex structure moduli space freely,\footnote{We have
  to take care of marginal stability walls along which the background
  ``gauge'' D-branes might decay, but generically they will not forbid
  all misalignment directions.} any brane wrapping a homology cycle
$[\beta]\neq [\alpha]$ can be misaligned from $\Sigma$. In particular,
our instanton can be misaligned from any brane not in its homology
class, and thus cannot have its $\ov\tau$ zero modes lifted by that
brane (and thus, no superpotential can be generated). This leaves the
cases where the brane wraps the same cycle as the instanton. The brane
could be exactly on top of the instanton (the gauge and $U(1)$ cases
described in the introduction) or it could be in a different cycle in
the same homology class. We present an example of this last and
somewhat more exotic case in the next section. An intriguing
possibility, explored in \cite{Heckman:2008es}, is to take the
instanton and the brane to wrap cycles which are different in local
homology, but are globally the same. We believe that in this case the
effect of the background $B$-field should also be taken into account.

\section{Homologically gauge instantons}

Let us shortly review a different kind of instanton effect in type II
which \emph{does} contribute to the superpotential, introduced in
\cite{Cvetic:2009mt}.

Our starting point is $\cN=4$ super-Yang mills. Instantons on top of
the gauge brane have all their zero modes lifted due to interactions
with the background gauge brane. In addition, they have the property
that their moduli space consists of two components, a Higgs branch on
which the instanton is dissolved into the gauge brane, and a Coulomb
branch in which the instanton can be taken away from the brane. It is
a simple calculation to show that even in the Coulomb branch the
$\ov\tau$ modes are lifted by interactions. This can be shown in
detail when the separation in the Coulomb branch is much smaller than
the string length, in this case the relevant zero mode lifting comes
from integrating out massive open strings between the instanton and
the brane. Denoting the separation of the gauge brane and the
instanton in the 6 internal directions as $\chi_m$ we obtain the
following effective coupling in the instanton action lifting the
$\ov\tau$ modes:
\begin{equation}
  e^{-S_{inst}(\chi_m)} = 8\pi^2g_0^{-3} (\ov\tau)^8\left(32 t^4 +
    144 t^2 + 64 - (32t^5 + 160 t^3 +
    120t)\sqrt{\pi}e^{t^2}\mathrm{erfc}(t)\right)
\end{equation}
where we have introduced the adimensional variable
$t\equiv\chi^2/\sqrt{g_0}$ (with $g_0^2 = 4\pi(4\pi^2\alpha')^{-2}
g_s$), and $\mathrm{erfc}(t)$ denotes the complementary error function
\cite{abramowitz+stegun}.

When the instanton and the brane are away from each other the dominant
channel is massless closed string exchange at tree level between the
instanton and the brane, with some insertions of $\ov\tau$ on the
instanton boundary. The requisite diagrams have been computed in the
context of M(atrix) theory \cite{Millar:2000ib} (for the coupling to
the RR four-form) and early D(-1) instanton studies
\cite{Green:1997tv} (for the coupling to the metric), and they are
non-vanishing:
\begin{equation}
  \begin{split}
  S_{inst} &\sim \ldots + \partial_m \partial_n G_{\rho\sigma}
  (\bar\Theta\Gamma^{m\rho\nu}\Theta)(\bar\Theta\Gamma^{n\sigma\nu}\Theta)\\
  S_{inst} &\sim \ldots + \partial_m \partial_n C^{(4)}_{\mu\nu\rho\sigma}
  (\bar\Theta\Gamma^{m\mu\nu}\Theta)(\bar\Theta\Gamma^{n\rho\sigma}\Theta)
\end{split}
\end{equation}
with $\Theta$ the spinor in the instanton worldvolume, which contains
$\ov\tau$.

So far we have only dealt with $\cN=4$, but the discussion will apply
with little change to any instanton with a Coulomb branch. A
particularly interesting class of examples comes from $\cN=2$
SYM. These theories can be engineered as D5 branes wrapping the blown
up cycle of an $A_1$ singularity:
\begin{equation}
  x^2+y^2+w^2 = 0.
\end{equation}
The interesting point is that this geometry can be easily fibered over
the remaining one complex dimensional direction in order to obtain an
$\cN=1$ compactification with isolated spheres:
\begin{equation}
  x^2+y^2+w^2 = f(z)^2.
\end{equation}
From the perspective of the worldvolume of the instanton, switching on
$f(z)$ corresponds to lifting the Coulomb branch, leaving only a set
of isolated points (the roots of $f(z)$). Since the original $\cN=2$
configuration lifted the $\ov\tau$ modes everywhere in the Coulomb
branch, we expect the $\ov\tau$ modes to be lifted on the deformed
$\cN=1$ geometry too, and thus to obtain contributions to the
superpotential.

\section{Lift of $O(1)$ instantons to F-theory}

In this section we will make some comments on how some known facts
about D-brane instantons lift to F-theory. We will derive results
which are already well understood from a CFT point of view in weakly
coupled type II, but we do it from a purely target space point of
view. In this way we will learn some lessons that carry over to
situations where CFT techniques are not available.

In particular, we will discuss the F-theory lift of $O(1)$ instantons,
and in particular how the orientifold removes the extra $\ov\tau$
fermionic zero modes.\footnote{As we will see below, the configuration
  that we study becomes a smooth F-theory model after quantum
  corrections, so Witten's arithmetic genus criterion
  \cite{Witten:1996bn} can be applied to the generic $O(1)$
  instanton. We will choose instead to work more
  microscopically. While the analysis is more involved in this way, it
  will illuminate some aspects of the instanton worldvolume theory
  which are not easy to see by looking just at the symmetries.}  The
discussion is simplest in the flat space setting, so let us consider a
F-theory compactification with base $\bR^{6}\times T^4$. We will
denote the coordinates $x^0,\ldots,x^9$ (with $x^6,\ldots,x^9$
parameterizing the $T^4$ torus). We want to lift the simplest $O(1)$
instanton, engineered as an instanton mapped to itself by the
orientifold action. Let us choose the orientifold to be of the $O7^-$
type (we choose conventions such that a D7 on top of the orientifold
has gauge group $O(1)$). This lifts in F-theory to a real codimension
2 degeneration of the fiber at $x^8=x^9 = 0$ (more precisely to a
couple of degenerations, see below). The instanton with gauge group
$O(1)$ will then be an euclidean D3 brane\footnote{We choose to work
  in a language close to type IIB. In the M-theory description we
  would have a M5 wrapping $T^4\times T^2$. After compactification of
  the 6 dimensional theory on the worldvolume of the M5 down to 4
  dimensions (by compactification on the $T^2$ fiber), the discussion
  below would also apply to the M-theory setting.} wrapping the $T^4$.

We would like to argue from a target space viewpoint that the
orientifold projects out the $\ov\tau$ modes present in the absence of
the $O7^-$. Consider the monodromy around the $O7^-$, it is given by
the following $SL(2,\bZ)$ element, up to conjugation:
\begin{equation}
  M_O = \begin{pmatrix}
    -1 & 4\\ 0 & -1
  \end{pmatrix},
  \label{eq:MO}
\end{equation}
where the 4 encodes the RR charge of the $O7^-$. The diagonal $-1$
part is the F-theory lift of the $(-1)^{F_L}\cdot \Omega$ orientifold
action in type IIB \cite{Sen:1996vd}, and will thus be responsible for
projecting out the $\ov\tau$ modes from the F-theory perspective, as
we show below.

The monodromy~\eqref{eq:MO} encodes the action of S-duality in the IIB
picture, and induces a S-duality action on the euclidean D3
worldvolume. Let us describe this worldvolume theory in some
detail. Before introducing the orientifold, we would have euclidean
$\cN=4$ $U(1)$ Yang-Mills theory living in a $T^4$. From the point of
view of this theory, introducing the orientifold amounts to
introducing a surface defect operator with the property that as we go
around it the fields in the $\cN=4$ theory enjoy a monodromy given by
$M_O$. Furthermore, the defect will break supersymmetry down to
$\cN=2$ (as we can easily see from the stringy description), and it
makes the theory non-conformal. We will discuss in detail the origin
and consequences of this non-conformality below.

It is hard to give a detailed description of the euclidean D3 brane
worldvolume theory with surface defects, however these simple
properties described above will suffice to analyze the effect of the
orientifold on zero modes. First of all, from \cite{Kapustin:2006pk}
for example, we learn that $M_O$ acts on the supersymmetries
$Q_{\alpha},\tilde Q_{\dot\alpha}$ of $\cN=4$ SYM in a chiral way
(here, and in the following, we will keep the $SU(4)_R$ spinor index
on the $\cN=4$ spinors implicit). Writing the general monodromy matrix
as
\begin{equation}
  M = \begin{pmatrix}
    a & b\\c & d
    \end{pmatrix}\,,
\end{equation}
the action of $M$ on supersymmetry charges is of the form
\begin{equation}
  Q_{\alpha}\to e^{i\varphi(M)}Q_{\alpha},\qquad \tilde Q_{\dot\alpha}\to
  e^{-i\varphi(M)}\tilde Q_{\dot\alpha}\,\,,
\end{equation}
with\footnote{The choice of notation is standard but somewhat
  unfortunate. $\ov\tau$ refers to the fermionic zero modes of the
  instanton, while $\tau$ denotes the complexified gauge coupling of
  $\cN=4$ super-Yang-Mills.}
\begin{equation}
  e^{i\varphi(M)}=\left(\frac{|c\tau + d|}{c\tau + d}\right)^{1/2}\,\,.
\end{equation}
In our case, this gives
\begin{equation}
  \label{eq:sign}
  Q_{\alpha}\to \sqrt{-1}Q_{\alpha},\qquad \tilde Q_{\dot\alpha}\to
  -\sqrt{-1}\tilde Q_{\dot\alpha}\,\,.
\end{equation}
Notice that this rotation is chiral, acting differently on right
handed and left handed fermionic supercharges. In the case of $\cN=4$
eq.~\eqref{eq:sign} is actually a harmless phase redefinition, since
it is contained in the center of the (anomaly-free) $SU(4)$ R-symmetry
group.

We have given the action of the orientifold S-duality on the
supersymmetries. The $\ov\tau,\theta$ modes that we are interested in
are in this language the gauginos of the $\cN=4$ theory. They are
obtained by acting with the supersymmetric transformations on the
gauge bosons, and thus transform just as the supercharges:
\begin{equation}
  \label{eq:gaugino-sign}
  \ov\tau \to \sqrt{-1}\ov\tau, \qquad \theta\to -\sqrt{-1}\theta.
\end{equation}

We will now argue that the action~\eqref{eq:sign} is actually
anomalous in our $\cN=2$ setting, and thus physical (in the sense that
the associated monodromy projects out states). The key point is that
\eqref{eq:gaugino-sign} is no longer anomaly-free after the insertion
of defects. Naively this would be easy to argue, since the rotation
that undoes \eqref{eq:gaugino-sign} is part of the $U(1)_R$ classical
symmetry of $\cN=2$ gauge theories, which is in the same multiplet as
the stress-energy tensor, and thus is anomalous if the theory is
non-conformal (which our theory is).

However, the analysis is more involved. The gauge instantons that
break $U(1)_R$ leave a discrete subgroup invariant. In the case of
ordinary $SU(N)$ $\cN=2$ with flavors the anomaly-free subgroup always
contains the $\bZ_4$ action \eqref{eq:gaugino-sign}. So the last step
that we need to show is that in the presence of the orientifold
defect, this universal $\bZ_4$ subgroup is actually partially broken
to $\bZ_2$ by anomalies ($\bZ_2$ always survives, as it corresponds to
multiplying all fermions by $-1$).

\begin{figure}[ht]
  \includegraphics[width=0.6\textwidth]{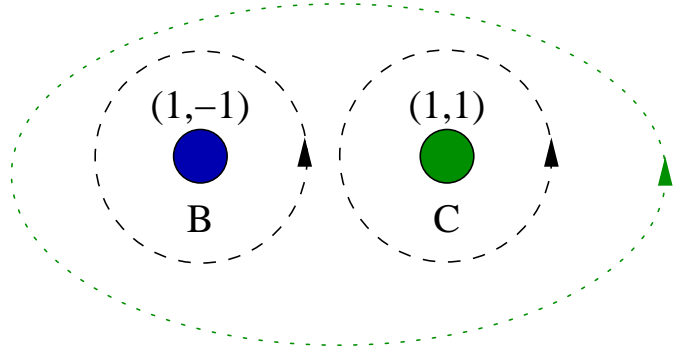}

  \caption{Schematic representation of the Euclidean D3-brane
    worldvolume theory after the orientifold splits, close to the
    orientifold locus. The two dots denote the location of the $B$ and
    $C$ defects. We have omitted the directions common to the $O7^-$
    and the Euclidean D3. The dotted green line denotes the contour
    along which we compute the monodromy. It factorizes into the two
    smaller dashed contours around each elementary degeneration.}

  \label{fig:E3}
\end{figure}

Let us present some evidence that this is actually the case. First of
all, let us use the fact that the $O7^-$ plane splits into two
separated degenerations once we lift it to F-theory
\cite{Sen:1996vd,Seiberg:1994rs} (we will refer to these two
degenerations as the \emph{elementary defects}).  In order to be
explicit, let us take the conventions of \cite{Gaberdiel:1998mv}. In
these conventions, the $O7^-$ plane splits into a $(1,-1)$ brane and a
$(1,1)$ brane. Let us denote them the $B$ and $C$ branes,
respectively.  See figure~\ref{fig:E3} for an illustration of the
system after the orientifold splits.

Recalling that the monodromy for a generic $(p,q)$ brane
is given (in a convenient $SL(2,\bZ)$ frame) by:
\begin{equation}
  M_{(p,q)} = \begin{pmatrix}
    1-pq & p^2 \\ -q^2 & 1+pq
    \end{pmatrix}  
\end{equation}
we have the following monodromies for the $B$ and $C$ branes:
\begin{equation}
  B=M_{(1,-1)} = \begin{pmatrix}
    2 & 1 \\ -1 & 0
  \end{pmatrix}\qquad
  C=M_{(1,1)} = \begin{pmatrix}
    0 & 1 \\ -1 & 2
  \end{pmatrix},
\end{equation}
where we have abused notation slightly, and denoted by $B$ the
monodromy matrix for $B$ branes, and similarly for $C$ branes. It is
easy to check that indeed $BC=M_O$, so the orientifold decomposes into
these two degenerations. We can now decompose the $U(1)_R$ rotation of
the supersymmetries due to the $SL(2,\bZ)$ transformation into a
couple of consecutive rotations, as we go around each degeneration separately:
\begin{equation}
  e^{i\varphi(O7^-)} = \sqrt{-1} = e^{i\varphi(M_B)} e^{i\varphi(M_C)}
\end{equation}
where
\begin{align}
  e^{i\varphi(M_C)} &= \sqrt{\frac{|-\tau+2|}{-\tau+2}}\\
  e^{i\varphi(M_B)} &= \sqrt{\frac{|\tau_C|}{-\tau_C}}.
\end{align}
We have introduced $\tau_C=1/(-\tau+2)$, the value of the coupling
after going around $C$. From these expressions we see that we can undo
the $U(1)_R$ rotation in two steps: first undo the rotation due to
$B$, and then the rotation due to $C$ (or vice versa). The overall
rotation will be anomalous if the addition of the induced shifts to
the $\cN=2$ $\theta$ angle is not a multiple of $2\pi$.\footnote{Here
  we are assuming that the anomaly we get when undoing the $U(1)_R$
  rotation due to each elementary defect comes only from zero modes
  localized at that defect. A possible justification for this
  assumption comes from the fact that the two defects are mutually
  non-local, and only the zero modes corresponding to one of them (in
  a particular frame where it is of type $(1,0)$) should be considered
  dynamical. This is consistent with the fact that in the F-theory
  lift there is a single complex structure modulus controlling the
  position of both elementary degenerations, once we fix the string
  coupling \cite{Sen:1996vd}.}

From the expressions above, for generic coupling, we see that it is
enough that the shift of the $\cN=2$ $\theta$ angle is not of the form
$4\times \varphi$ (here $\varphi$ denotes the $U(1)_R$ rotation
angle), as it would be for usual $\cN=2$ theory with flavors, but
$2\times\varphi$. This is actually a plausible result, for the
following reason. The factor of 4 comes from the number of zero modes
in the one instanton background in which we compute the anomaly. The
background in which we want to compute our anomaly is not a
one-instanton background, but rather a one-vortex background. There is
a deep relation between vortices and instantons (see for example
\cite{Tong:2005un}), and in particular the dimension of the
$k$-instanton moduli space is $4kN_c$, while the dimension of the
$k$-vortex moduli space is $2kN_c$ (in our case we have $N_c = k =
1$). In situations with enough supersymmetry, the bosonic dimension of
the moduli space agrees with the fermionic dimension of the moduli
space, and thus we expect that the number of fermionic zero modes of a
vortex is exactly half the number of fermionic zero modes of an
instanton, leading to half a contribution to the anomaly.

Based on this evidence (and from our knowledge of the microscopic
picture) this leads us to argue that the correct shift of the axion in
the vortex background is indeed $2\varphi$, and thus the action of
$SL(2,\bZ)$ cannot be undone by an anomaly-free $U(1)_R$ rotation.

Equipped with these arguments, we see that upon circling once around
the defect, the $\cN=4$ field theory gets mapped to itself, except for
the chiral rotation of the gauginos. We can multiply all fermions by
$\sqrt{-1}$ (a harmless phase redefinition\footnote{At this point we
  could have also chosen $-\sqrt{-1}$. The sign is determined by
  imposing that the resulting projection coming
  from~\eqref{eq:final-sign} is the one compatible with the
  supersymmetries preserved by the instanton-$O7^-$ system.}), and we
obtain the following monodromy action:
\begin{equation}
  \label{eq:final-sign}
  \theta\to \theta, \qquad \ov\tau \to -\ov\tau.
\end{equation}
Notice that we are mapping the gauge coupling to itself: $\tau\to
\tau-4$ (the integer shift in the theta angle has no effect on the
theory), and in particular far away from the orientifold we are
mapping weak coupling to weak coupling, so we can identify degrees of
freedom straightforwardly. The minus sign in front of $\ov\tau$ then
tells us that this mode is unphysical, and projected out by the
orientifold.

Note also that the crucial point was the $\sqrt{-1}$ component of the
orientifold monodromy action \eqref{eq:sign} on the fermions, which
does not act on the gauge coupling. This $\sqrt{-1}$ factor is absent
in the case of pure D7 branes, where we have
$e^{i\varphi(M)}=\pm\sqrt{1}$, giving a non-chiral (and thus
anomaly-free) phase rotation. In fact this is precisely the
anomaly-free $\bZ_2$ subgroup that was preserved by the orientifold.


\begin{theacknowledgments}
  We would like to acknowledge interesting discussions with Ralph
  Blumenhagen, Jim Halverson, Liam McAllister, Tony Pantev, Mike
  Schulz, Angel Uranga, and Timo Weigand. We would also like to thank
  Jonathan Heckman, Joseph Marsano, Natalia Saulina, Sakura
  Sch{\"a}fer-Nameki and Cumrun Vafa for extensive discussions after
  \cite{Cvetic:2009mt} appeared. I.G.E. thanks Nao Hasegawa for kind
  support and constant encouragement. This work is supported by DOE
  grant DE-FG05-95ER40893-A020, NSF RTG grant DMS-0636606 and Fay
  R. and Eugene L. Langberg Chair.
\end{theacknowledgments}



\bibliographystyle{aipproc}   

\bibliography{refs}

\begin{thebibliography}{53}
\expandafter\ifx\csname natexlab\endcsname\relax\def\natexlab#1{#1}\fi
\providecommand{\enquote}[1]{``#1''}
\expandafter\ifx\csname url\endcsname\relax
  \def\url#1{\texttt{#1}}\fi
\expandafter\ifx\csname urlprefix\endcsname\relax\def\urlprefix{URL }\fi
\providecommand{\eprint}[2][]{\url{#2}}

\bibitem[Becker et~al.(1995)]{Becker:1995kb}
K.~Becker, M.~Becker, and A.~Strominger, \emph{Nucl. Phys.} \textbf{B456},
  130--152 (1995), \eprint{hep-th/9507158}.

\bibitem[Blumenhagen et~al.(2007{\natexlab{a}})]{Blumenhagen:2006xt}
R.~Blumenhagen, M.~Cveti\v{c}, and T.~Weigand, \emph{Nucl. Phys.}
  \textbf{B771}, 113--142 (2007{\natexlab{a}}), \eprint{hep-th/0609191}.

\bibitem[Ib{\'a}{\~n}ez and Uranga(2007)]{Ibanez:2006da}
L.~E. Ib{\'a}{\~n}ez, and A.~M. Uranga, \emph{JHEP} \textbf{03}, 052 (2007),
  \eprint{hep-th/0609213}.

\bibitem[Florea et~al.(2007)]{Florea:2006si}
B.~Florea, S.~Kachru, J.~McGreevy, and N.~Saulina, \emph{JHEP} \textbf{05}, 024
  (2007), \eprint{hep-th/0610003}.

\bibitem[Blumenhagen et~al.(2009)]{Blumenhagen:2009qh}
R.~Blumenhagen, M.~Cveti{\v c}, S.~Kachru, and T.~Weigand  (2009),
  \eprint{0902.3251}.

\bibitem[Cveti{\v{c}} et~al.(2009)]{Cvetic:2009mt}
M.~Cveti{\v{c}}, I.~Garc{\'i}a-Etxebarria, and R.~Richter  (2009),
  \eprint{0905.1694}.

\bibitem[Witten(1996{\natexlab{a}})]{Witten:1995gx}
E.~Witten, \emph{Nucl. Phys.} \textbf{B460}, 541--559 (1996{\natexlab{a}}),
  \eprint{hep-th/9511030}.

\bibitem[Douglas(1998)]{Douglas:1996uz}
M.~R. Douglas, \emph{J. Geom. Phys.} \textbf{28}, 255--262 (1998),
  \eprint{hep-th/9604198}.

\bibitem[Bill{\`o} et~al.(2003)]{Billo:2002hm}
M.~Bill{\`o}, et~al., \emph{JHEP} \textbf{02}, 045 (2003),
  \eprint{hep-th/0211250}.

\bibitem[Akerblom et~al.(2007)]{Akerblom:2006hx}
N.~Akerblom, R.~Blumenhagen, D.~L{\"u}st, E.~Plauschinn, and
  M.~Schmidt-Sommerfeld, \emph{JHEP} \textbf{04}, 076 (2007),
  \eprint{hep-th/0612132}.

\bibitem[Bianchi and Kiritsis(2007)]{Bianchi:2007fx}
M.~Bianchi, and E.~Kiritsis, \emph{Nucl. Phys.} \textbf{B782}, 26--50 (2007),
  \eprint{hep-th/0702015}.

\bibitem[Argurio et~al.(2007{\natexlab{a}})]{Argurio:2007vqa}
R.~Argurio, M.~Bertolini, G.~Ferretti, A.~Lerda, and C.~Petersson, \emph{JHEP}
  \textbf{06}, 067 (2007{\natexlab{a}}), \eprint{0704.0262}.

\bibitem[Bianchi et~al.(2007)]{Bianchi:2007wy}
M.~Bianchi, F.~Fucito, and J.~F. Morales, \emph{JHEP} \textbf{07}, 038 (2007),
  \eprint{0704.0784}.

\bibitem[Aganagic et~al.(2008)]{Aganagic:2007py}
M.~Aganagic, C.~Beem, and S.~Kachru, \emph{Nucl. Phys.} \textbf{B796}, 1--24
  (2008), \eprint{0709.4277}.

\bibitem[Garc{\'i}a-Etxebarria and Uranga(2008)]{GarciaEtxebarria:2007zv}
I.~Garc{\'i}a-Etxebarria, and A.~M. Uranga, \emph{JHEP} \textbf{01}, 033
  (2008), \eprint{0711.1430}.

\bibitem[Petersson(2008)]{Petersson:2007sc}
C.~Petersson, \emph{JHEP} \textbf{05}, 078 (2008), \eprint{0711.1837}.

\bibitem[Garc{\'i}a-Etxebarria(2009)]{GarciaEtxebarria:2008iw}
I.~Garc{\'i}a-Etxebarria, \emph{JHEP} \textbf{07}, 017 (2009),
  \eprint{0810.1482}.

\bibitem[Ferretti and Petersson(2009)]{Ferretti:2009tz}
G.~Ferretti, and C.~Petersson, \emph{JHEP} \textbf{03}, 040 (2009),
  \eprint{0901.1182}.

\bibitem[Argurio et~al.(2007{\natexlab{b}})]{Argurio:2007qk}
R.~Argurio, M.~Bertolini, S.~Franco, and S.~Kachru, \emph{JHEP} \textbf{06},
  017 (2007{\natexlab{b}}), \eprint{hep-th/0703236}.

\bibitem[Ib{\'a}{\~n}ez et~al.(2007)]{Ibanez:2007rs}
L.~E. Ib{\'a}{\~n}ez, A.~N. Schellekens, and A.~M. Uranga, \emph{JHEP}
  \textbf{06}, 011 (2007), \eprint{0704.1079}.

\bibitem[Marolf et~al.(2003{\natexlab{a}})]{Marolf:2003ye}
D.~Marolf, L.~Martucci, and P.~J. Silva, \emph{JHEP} \textbf{04}, 051
  (2003{\natexlab{a}}), \eprint{hep-th/0303209}.

\bibitem[Marolf et~al.(2003{\natexlab{b}})]{Marolf:2003vf}
D.~Marolf, L.~Martucci, and P.~J. Silva, \emph{JHEP} \textbf{07}, 019
  (2003{\natexlab{b}}), \eprint{hep-th/0306066}.

\bibitem[Kallosh et~al.(2005)]{Kallosh:2005gs}
R.~Kallosh, A.-K. Kashani-Poor, and A.~Tomasiello, \emph{JHEP} \textbf{06}, 069
  (2005), \eprint{hep-th/0503138}.

\bibitem[Martucci et~al.(2005)]{Martucci:2005rb}
L.~Martucci, J.~Rosseel, D.~Van~den Bleeken, and A.~Van~Proeyen, \emph{Class.
  Quant. Grav.} \textbf{22}, 2745--2764 (2005), \eprint{hep-th/0504041}.

\bibitem[Bergshoeff et~al.(2005)]{Bergshoeff:2005yp}
E.~Bergshoeff, R.~Kallosh, A.-K. Kashani-Poor, D.~Sorokin, and A.~Tomasiello,
  \emph{JHEP} \textbf{10}, 102 (2005), \eprint{hep-th/0507069}.

\bibitem[Park(2005)]{Park:2005hj}
J.~Park  (2005), \eprint{hep-th/0507091}.

\bibitem[Bandos and de~Azcarraga(2005)]{Bandos:2005ww}
I.~Bandos, and J.~A. de~Azcarraga, \emph{JHEP} \textbf{09}, 064 (2005),
  \eprint{hep-th/0507197}.

\bibitem[L{\" u}st et~al.(2006)]{Lust:2005cu}
D.~L{\" u}st, S.~Reffert, W.~Schulgin, and P.~K. Tripathy, \emph{JHEP}
  \textbf{08}, 071 (2006), \eprint{hep-th/0509082}.

\bibitem[Tsimpis(2007)]{Tsimpis:2007sx}
D.~Tsimpis, \emph{JHEP} \textbf{03}, 099 (2007), \eprint{hep-th/0701287}.

\bibitem[Schulgin(2007)]{Schulgin:2007zz}
W.~Schulgin, \emph{Nucl. Phys. Proc. Suppl.} \textbf{171}, 316--318 (2007).

\bibitem[Blumenhagen et~al.(2007{\natexlab{b}})]{Blumenhagen:2007bn}
R.~Blumenhagen, M.~Cveti\v{c}, R.~Richter, and T.~Weigand, \emph{JHEP}
  \textbf{10}, 098 (2007{\natexlab{b}}), \eprint{0708.0403}.

\bibitem[Garc{\'i}a-Etxebarria et~al.(2008)]{GarciaEtxebarria:2008pi}
I.~Garc{\'i}a-Etxebarria, F.~Marchesano, and A.~M. Uranga, \emph{JHEP}
  \textbf{07}, 028 (2008), \eprint{0805.0713}.

\bibitem[Billo' et~al.(2008{\natexlab{a}})]{Billo':2008pg}
M.~Billo', et~al., \emph{JHEP} \textbf{12}, 102 (2008{\natexlab{a}}),
  \eprint{0807.4098}.

\bibitem[Billo' et~al.(2008{\natexlab{b}})]{Billo':2008sp}
M.~Billo', et~al., \emph{JHEP} \textbf{10}, 112 (2008{\natexlab{b}}),
  \eprint{0807.1666}.

\bibitem[Uranga(2009)]{Uranga:2008nh}
A.~M. Uranga, \emph{JHEP} \textbf{01}, 048 (2009), \eprint{0808.2918}.

\bibitem[Heckman et~al.(2008)]{Heckman:2008es}
J.~J. Heckman, J.~Marsano, N.~Saulina, S.~Sch{\"a}fer-Nameki, and C.~Vafa
  (2008), \eprint{0808.1286}.

\bibitem[Marsano et~al.(2008)]{Marsano:2008py}
J.~Marsano, N.~Saulina, and S.~Sch{\"a}fer-Nameki  (2008), \eprint{0808.2450}.

\bibitem[Banks et~al.(1988)]{Banks:1987cy}
T.~Banks, L.~J. Dixon, D.~Friedan, and E.~J. Martinec, \emph{Nucl. Phys.}
  \textbf{B299}, 613--626 (1988).

\bibitem[Banks and Dixon(1988)]{Banks:1988yz}
T.~Banks, and L.~J. Dixon, \emph{Nucl. Phys.} \textbf{B307}, 93--108 (1988).

\bibitem[Cveti{\v c} et~al.(2007)]{Cvetic:2007ku}
M.~Cveti{\v c}, R.~Richter, and T.~Weigand, \emph{Phys. Rev.} \textbf{D76},
  086002 (2007), \eprint{hep-th/0703028}.

\bibitem[Green and Gutperle(2000)]{Green:2000ke}
M.~B. Green, and M.~Gutperle, \emph{JHEP} \textbf{02}, 014 (2000),
  \eprint{hep-th/0002011}.

\bibitem[Beasley and Witten(2005)]{Beasley:2004ys}
C.~Beasley, and E.~Witten, \emph{JHEP} \textbf{01}, 056 (2005),
  \eprint{hep-th/0409149}.

\bibitem[Beasley and Witten(2006)]{Beasley:2005iu}
C.~Beasley, and E.~Witten, \emph{JHEP} \textbf{02}, 060 (2006),
  \eprint{hep-th/0512039}.

\bibitem[Aspinwall(2004)]{Aspinwall:2004jr}
P.~S. Aspinwall  (2004), \eprint{hep-th/0403166}.

\bibitem[Abramowitz and Stegun(1964)]{abramowitz+stegun}
M.~Abramowitz, and I.~A. Stegun, Handbook of mathematical functions with
  formulas, graphs, and mathematical tables,
  \url{http://www.math.sfu.ca/~cbm/aands/} (1964).

\bibitem[Millar et~al.(2000)]{Millar:2000ib}
K.~Millar, W.~Taylor, and M.~Van~Raamsdonk  (2000), \eprint{hep-th/0007157}.

\bibitem[Green and Gutperle(1997)]{Green:1997tv}
M.~B. Green, and M.~Gutperle, \emph{Nucl. Phys.} \textbf{B498}, 195--227
  (1997), \eprint{hep-th/9701093}.

\bibitem[Witten(1996{\natexlab{b}})]{Witten:1996bn}
E.~Witten, \emph{Nucl. Phys.} \textbf{B474}, 343--360 (1996{\natexlab{b}}),
  \eprint{hep-th/9604030}.

\bibitem[Sen(1996)]{Sen:1996vd}
A.~Sen, \emph{Nucl. Phys.} \textbf{B475}, 562--578 (1996),
  \eprint{hep-th/9605150}.

\bibitem[Kapustin and Witten(2006)]{Kapustin:2006pk}
A.~Kapustin, and E.~Witten  (2006), \eprint{hep-th/0604151}.

\bibitem[Seiberg and Witten(1994)]{Seiberg:1994rs}
N.~Seiberg, and E.~Witten, \emph{Nucl. Phys.} \textbf{B426}, 19--52 (1994),
  \eprint{hep-th/9407087}.

\bibitem[Gaberdiel et~al.(1998)]{Gaberdiel:1998mv}
M.~R. Gaberdiel, T.~Hauer, and B.~Zwiebach, \emph{Nucl. Phys.} \textbf{B525},
  117--145 (1998), \eprint{hep-th/9801205}.

\bibitem[Tong(2005)]{Tong:2005un}
D.~Tong  (2005), \eprint{hep-th/0509216}.

\end{thebibliography}

\end{document}